\documentclass{phbauth}       
\usepackage{graphicx}

\begin{document}
\newcommand{\M}{\mathrm}

\begin{frontmatter}

\title{Quenched Kosterlitz-Thouless superfluid transitions}

\author{Han-Ching Chu},
\author{Gary A. Williams\thanksref{thank1}}
\address{Department of Physics and Astronomy, University of 
California, Los Angeles, CA 90095, USA}

\thanks[thank1]{Corresponding author. E-mail: gaw@ucla.edu} 

\begin{abstract}
The properties of rapidly quenched superfluid phase transitions are
computed for two-dimensional Kosterlitz-Thouless (KT) systems.  The
decay in the vortex-pair density and the recovery of the superfluid
density after a quench are found by solving the Fokker-Planck equation 
describing the vortex dynamics, in conjunction with the KT recursion 
relations.  The vortex density is found to decay approximately as the 
inverse of the time from the quench, in agreement with computer 
simulations and with scaling theories. 
\end{abstract}


\end{frontmatter}
The properties of quenched phase transitions are of interest
because they may be relevant to the rapidly cooling early 
universe \cite{antunes}.  
The cosmic-string phase transitions in that case have recently been 
linked with the 
vortex-loop superfluid transition in liquid helium and high-T$_{c}$ 
superconductors \cite{prl}.  In this paper we investigate a new 
technique for calculating the quenched superfluid transition in two 
dimensions, employing Kosterlitz-Thouless renormalization methods.
The results are in agreement with previous studies that used either 
computer simulations \cite{sim} or scaling 
theories \cite{bray}.  

The stochastic dynamics of the vortex pairs of the 
Kosterlitz-Thouless theory are modeled by a Fokker-Planck 
equation \cite{ahns}
for the distribution function $\Gamma (r,t)$, which is the density of 
vortex pairs of separation $r$\,:
\begin{equation}
{{\partial \,\Gamma } \over {\partial \,t'}}=\,\;{\partial  \over 
{\partial \kern 1pt l}}\cdot \left( {{{\partial \Gamma } \over 
{\,\partial \kern 1pt l}}+{\Gamma  \over {k_BT}}{{\partial \kern 1pt U} 
\over {\,\partial \kern 1pt l}}} \right) \quad ,
\end{equation}
where $l=\ln ({r \mathord{\left/ {\vphantom {r {a_o)}}} 
\right. \kern-\nulldelimiterspace} {a_o)}}$, with $a_{o}$ the vortex 
core radius, and $U$ is the interaction potential between a pair.  
The time here is in units of the diffusion time of 
the smallest pairs of separation $a_{o}$, $t' = t/\tau_{o}$, where 
$\tau _o^{-1}= {2D}/{a_o^2}$, with $D$ the diffusion 
coefficient. From the KT theory
\begin{equation}
{1 \over {k_BT}}{{\partial U} \over {\partial l}}=2\pi \,K \quad,
\end{equation}
where $K={\hbar ^2\sigma _s}$/${m^2k_BT}$ is the dimensionless areal 
superfluid density.  $K$ is renormalized by the presence of the vortex 
pairs, and the KT recursion relation for $K$ can be written in terms 
of $\Gamma$ as
\begin{equation}
{{\partial K} \over {\partial \kern 1pt l}}
=-4\pi ^3\,K^2\,\Gamma \,{\M e}^{4l} \quad .
\end{equation}

\begin{figure}[ht]
\begin{center}\leavevmode
\includegraphics[width=1.0\linewidth]{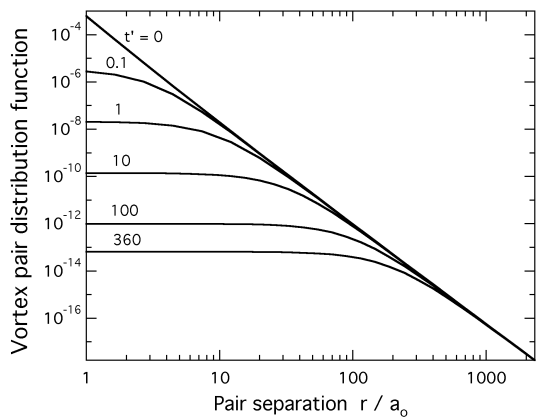}
\caption{Distribution function $\Gamma$ as a function of the pair separation
for different times after the quench.}\label{figure1}\end{center}\end{figure}

\begin{figure}[ht]
\begin{center}\leavevmode
\includegraphics[width=1.0\linewidth]{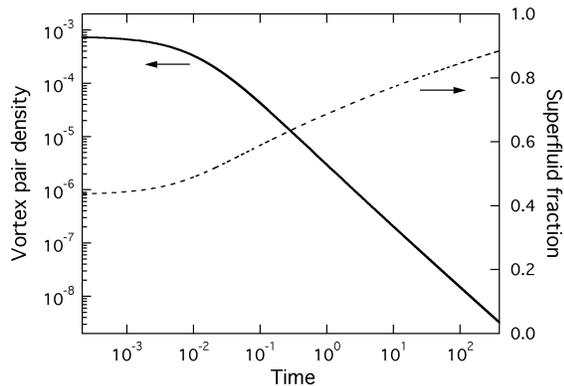}
\caption{Vortex density and superfluid fraction as a function of time.}
\label{figure2}\end{center}\end{figure}

To quench from $T_{\rm KT}$ to low temperature it is first
necessary to compute $K(l)$ and the vortex distribution at
$T_{\rm KT}$.   The equilibrium solution of Eq.\ (1) is just the
Boltzmann distribution,  and taking the normalized core energy of a
pair to be
$\pi^{2} K(0)$, iteration of the KT recursion relations gives the
$t'=0$ vortex distribution  shown in Fig. 1.  The iterations are taken
to a maximum pair separation of 
$l=8$ ($r/a_{o}$ = 3000).

We then (instantaneously) change the temperature to $0.1 \,T_{\rm KT}$ 
($K(0) \rightarrow 10\,K(0)$) and 
compute the new renormalized values of $K(l)$ from Eq.\ (3), using a 
Runge-Kutta method.  Eq.\ (1) is then solved for the changed $\Gamma$ 
after a time step $\Delta t'$ with a two-step Lax-Wendroff algorithm.
The changed vortex distribution is used to recompute $K(l)$ 
from Eq.\ (3) at the increased time, and the entire sequence is 
continuously repeated to step out in time. 
In quenching to a low temperature the second term in the parentheses 
of the right side of Eq.\ (1) is the dominant term, and as a first 
approximation we have neglected the first term.

The results for the pair distribution function (in units $a_{o}^{-4}$)
are shown in Fig. 1.  This is a classic phase-ordering system, and as 
expected the smallest pairs decay away first, while the largest pairs 
remain ''frozen'' until the longest times. By integrating the 
distribution function the vortex density can be computed, shown in 
Fig. 2 (in units $a_{o}^{-2}$). At long times the density decays 
approximately as $t'^{-1}$, in agreement with previous simulations and 
scaling theories \cite{sim,bray}, although it appears to be 
approaching this quite slowly (the exponent is -1.17
at $t'=1$ and -1.12 at $t'=250$), probably because $K$ and 
the superfluid fraction ($K(8)/K(0)$, dashed curve in Fig. 2) are still 
changing fairly rapidly with time.  The $1/t'$
behavior can be traced to the nonlinear term proportional to 
$\Gamma ^{2}$ that results from the substitution of Eqs.\ (2) and (3)
into Eq.\ (1). 

A key feature of the present results is the scaling of the vortex 
decay with the diffusion time $\tau_{o}$, something generally 
neglected in previous treatments \cite{antunes,sim,bray}.  Estimating
for helium films $D \sim 1\cdot 10^{-4}$ cm$^{2}$/s and 
$a_{o}\sim 1\cdot 10^{-8}$ cm gives $\tau_{o}\sim
1\cdot 10^{-12}$ s,  in agreement with 
estimates of the same quantity in three dimensions \cite{dyn}.  The 
very fast decay times may well explain the inability of millisecond-scale
experiments to detect the vortices \cite{mcclintock}.

This work is supported by the U. S. National Science Foundation, 
DMR 97-31523.


\begin{thebibliography}{9}
    
\bibitem{antunes} N. Antunes, L. Bettencourt, W. Zurek, 
Phys. Rev. Lett. {82} (1999) 2824.

\bibitem{prl} G. A. Williams,
Phys. Rev. Lett. {82} (1999) 1201.

\bibitem{sim} M. Mondello, N. Goldenfeld,
Phys. Rev. A {42} (1990) 5865; B. Yurke, A. Pargellis,
T. Kovacs, D. Huse,
Phys. Rev. E {47} (1993) 1525.

\bibitem{bray} A. Rutenberg, A. Bray,
Phys. Rev. E {51} (1995) R1641.

\bibitem{ahns} V. Ambegaokar, B. Halperin, D. Nelson, E. Siggia, 
Phys. Rev. B {21} (1980) 1806.

\bibitem{dyn} G. A. Williams,
     J. Low Temp. Phys. 
	 {93} (1993) 1079.

\bibitem{mcclintock}
     M. Dodd {\it et al.},
     Phys. Rev. Lett.
     {81} (1998) 3703.


\end{thebibliography}
\end{document}